# A tutorial comparing different covariate balancing methods with an application evaluating the causal effects of substance use treatment programs for adolescents

Andreas Markoulidakis [a,c,*], Khadijeh Taiyari[c], *Peter Holmans*[b], Philip Pallmann[c],

Monica Busse, Mark D. Godley[e], Beth Ann Griffin[d]

[a]School of Medicine, Cardiff University, Cardiff, UK
[b]Division of Psychological Medicine and Clinical Neurosciences, School of Medicine, Cardiff University, Cardiff, UK
[c]Centre for Trials Research, Cardiff University, Cardiff, UK
[d]RAND Corporation, Arlington, VA, USA

[*]*MarkoulidakisA@cardiff.ac.uk*

Randomized controlled trials are the gold standard for measuring causal effects. However, they are often not always feasible, and causal treatment effects must be estimated from observational data. Observational studies do not allow robust conclusions about causal relationships unless statistical techniques account for the imbalance of pretreatment confounders across groups while key assumptions hold. Propensity score and balance weighting (PSBW) are useful techniques that aim to reduce the imbalances between treatment groups by weighting the groups to look alike on the observed confounders. There are many methods available to estimate PSBW. However, it is unclear a priori which will achieve the best trade-off between covariate balance and effective sample size. Moreover, it is critical to assess the validity of key assumptions required for robust estimation of the needed treatment effects, including the overlap and no unmeasured confounding assumptions. We present a step-by-step guide to covariate balancing strategies, including how to evaluate overlap, obtain estimates of PSBW, check for covariate balance, and assess sensitivity to unobserved confounding. We compare the performance of several estimation methods using a case study examining the relative effectiveness of substance use treatment programs and provide a user-friendly web application that can implement the proposed steps.

## Introduction

In a randomized controlled trial (RCT), the assignment to treatment or control group is done using randomization, to ensure that there is no systematic bias from observed and unobserved confounders when estimating the effect of the treatment. For sufficiently large sample sizes, the two groups will usually have similar baseline characteristics so that the groups are comparable to one another (Altman & Bland, 1999). Unfortunately, such is not the case with observational studies where randomization to treatment assignments is not possible. In observational studies, group assignment might be due to factors that the researcher cannot control, such as underlying conditions of the individual. These differences between the groups will introduce bias into the estimated effects of a treatment. For example, the choice to exercise among individuals with a disease might be influenced in part based on the severity of their disease: those with more severe disease might be less keen or able to exercise, and any subsequent comparison between those who did and did not exercise will be distorted, as it will compare groups with different pre-treatment levels of disease severity. Alternatively, in many health services research applications, there may not be funding available to test the effectiveness of promising evidence-based treatments in real world settings but there may be access to rich, secondary data sets (like electronic health records or insurance data sets) that follow individuals over time who self-select into different types of treatments or health services. Naturally, the group of individuals who self-select into treatments or health services will be notably different from those that do not in terms of both socio-demographic characteristics and other health indicators. Our motivating example is based on such a rich data source, the Global Appraisal of Individual Needs (GAIN) data base which has followed more than 20,000 adolescent substance users over time as they received access to and treatment with promising evidence-based treatments (Dennis et al., 2003) such as the Adolescent Community Reinforcement Approach (A-CRA) (Godley, 2001) and Motivational Enhancement Treatment/Cognitive Behavior Therapy, 5 Sessions (MET/CBT5) (Sampl & Kadden, 2001). Specifically, we will illustrate how to estimate the relative effectiveness of these two treatment programs for adolescents on total days abstinent 6-months later.

Estimation of accurate causal treatment effects (Holland, 1986) is the main goal of many observational studies. The causal effect of a treatment for each individual is defined as the difference in the outcome for an individual had they received that treatment compared to the outcome had they not received it. This is practically impossible to measure directly since typically, only one treatment condition will be assigned to each individual (Rosenbaum & Rubin, 1983). As a consequence, we design longitudinal, observational studies, where two (or more) groups receive different treatments over time, and then we estimate the causal treatment effect from the difference in the outcomes of the groups. Without adjustments for confounding, however, this estimate will almost certainly be biased.

The propensity score (PS) (Rosenbaum & Rubin, 1983) is the probability of an individual's allocation to a specific treatment group, given their observed baseline (pre-treatment) characteristics. The PS can be used to create comparable treatment groups by either weighting, matching, adjusting or stratifying on the PS. By minimising the imbalance of known and observed confounders between the treatment groups, PS methods reduce the bias in the estimation of the causal treatment effect due to the observed confounders. Here we consider only PS weighting (Hernán et al., 2000; Robins et al., 2000) (as opposed to PS matching, stratification and adjustment) given the notable increase in methods for estimating weights that have arisen (Elze et al., 2017; Harder et al., 2010; Olmos & Govindasamy, 2015; Posner & Ash, 2012). In addition to PS weights, we consider the closely related balancing weights via entropy balancing (Hainmueller, 2012). Entropy balancing computes the balancing weight directly, as opposed to traditional PS algorithms, which at first compute the PS and then transform it to weights. Thus, we are focusing only on PS and balancing weights methods to allow for a direct comparison of the performance of different algorithms.

There are several estimation methods for both PS and balancing weights, including parametric and non-parametric modeling and machine learning techniques. However, there is no clear indication that a single method performs best in every dataset (Griffin et al., 2017; Setodji et al., 2017; Setoguchi et al., 2008). This is the reason, we suggest, that one should consider multiple methods and finally estimate the causal treatment effect based on the one that best balances the treatment groups without reducing the effective sample size unduly. This tutorial will present an implementation of some of the more commonly used PS estimation methods, namely Logistic Regression (LR) (Agresti, 2018; Wright, 1995), Generalized Boosted Model (GBM) (McCaffrey et al., 2004), and Covariate Balance Propensity Score (CBPS) (Imai & Ratkovic, 2014). We will also consider the use of balancing weights via Entropy Balancing (EB) (Hainmueller, 2012) and provide a summary of the advantages and disadvantages of each approach. The performance of the methods will be illustrated by applying them to understanding the relative effects of A-CRA versus MET/CBT5 on days abstinent for adolescent substance users.

The remainder of the article is organised as follows. *Section The 6 Key Steps towards Estimating Causal Treatment Effects* summarizes the six key steps needed to estimate causal treatment effects of a treatment in observational studies. Then, the PS and EB estimation algorithms we use are described in *section Propensity Scoring and Balancing Weight Analysis Methods* , and the measures of performance of the different balancing methods are briefly discussed in *section Measures to Evaluate Balance* . *Section Study Data* describes the data example and *section Obtaining PS Weighting Estimates and Assessing their Performance* uses the data to walk through the 6 key steps needed to estimate causal effects. *Section Discussion/Conclusion* concludes. We have developed a user-friendly Shiny application that can be used to run all steps proposed in this tutorial on any data set.

## The 6 Key Steps towards Estimating Causal Treatment Effects

There is a wide discussion in the literature, regarding the number of steps necessary to estimate causal treatment effects using balancing and PS weights (Bergstra et al., 2019; Caliendo & Kopeinig, 2008; Setodji et al., 2017). Here, we follow 6 key steps, uniquely considering the relative performance of several estimation methods for the balancing and PS weights as well as demonstrating the needed and, often underutilized, use of omitted variable sensitivity analyses.

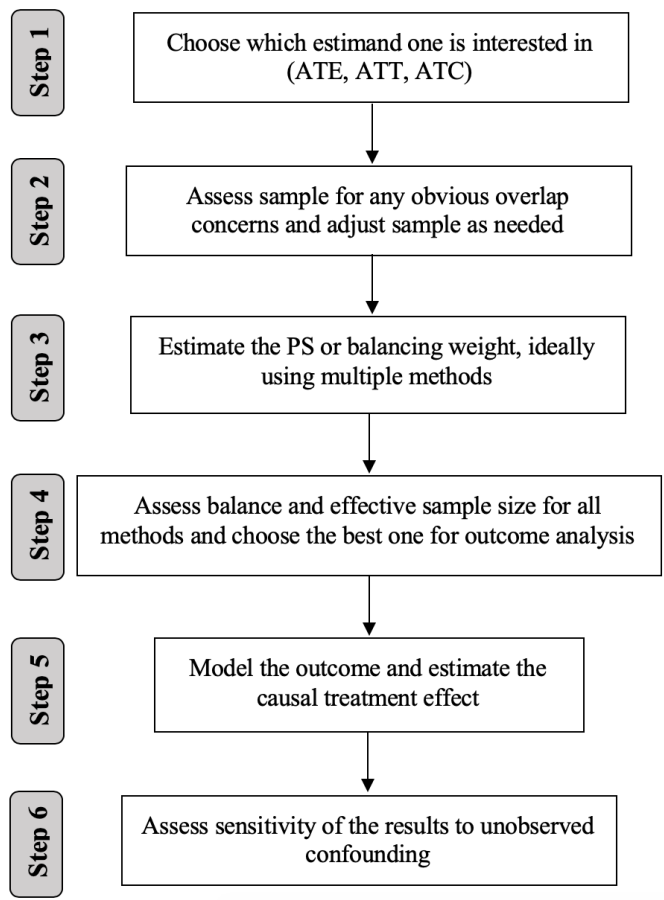

1. **Step 1.** *Choose which estimand one is interested in (ATE, ATT, ATC).*

   In order to define the commonly used causal treatment effects, we use the potential outcomes notation first introduced by (Rosenbaum & Rubin, 1983). For each individual in a study with two groups, we define $Y_{1i}$ to denote their potential outcome under treatment and $Y_{0i}$ to denote their potential outcome under control for $i = 1, \ldots, N$. While these potential outcomes exist in theory for all individuals in our study, we only get to observe one potential outcome for each individual. Namely, we observe the outcome for the treatment they actually ended up receiving. We define our treatment indicator as $T_i$ where values of 1 denote that individual received treatment and 0 denotes receipt of the control condition. Then we can define $Y_i^{obs} = Y_{1i} \cdot T_i + Y_{0i} \cdot (1 - T_i)$.

   The most commonly used causal treatment effects are:

   - the *Average Treatment Effect on the Entire population*

   $$ATE = E[Y_1] - E[Y_0],$$

   - the *Average Treatment Effect on the Treated population*

   $$ATT = E[Y_1|T = 1] - E[Y_0|T = 1],$$

- the *Average Treatment Effect on the Control population*

$$ATC = E[Y_1|T = 0] - E[Y_0|T = 0].$$

Each estimand is used for a different purpose, depending on the research question a study is trying to answer. ATE allows one to understand the average causal treatment effect for the entire population of individuals in both the treatment and control groups. In contrast, ATT allows one to understand the effect of the treatment among only individuals like those in the treatment group, and ATC to understand the effect of the treatment among only individuals like those in the control group.

For example, considering our case study, the ATE would measure the relative effect of the two treatments for all adolescents in our sample, while ATT and ATC would quantify the relative effect of the two treatments for adolescents like those in the individual treatment groups. If we assume A-CRA is officially to be defined as the "treatment" group and MET/CBT5 as the "control/comparison" condition, then ATT measures the relative effectiveness of the two programs for youth like those in the A-CRA group while ATC measures the relative effectiveness of the programs for youth like those in the MET/CBT5 group. Since adolescents in the two groups tend to be different in several ways (e.g., youth in the A-CRA group were older, more likely to be a minority and had higher mean levels of substance use and mental health problems and past substance use and mental health treatment than youth in the MET/CBT5 group), the populations to which ATT and ATC generalize will be different.

2. **Step 2.** *Assess sample for any obvious overlap concerns and adjust as needed.*

   Rubin's Causal Model (RCM) is the first widely known approach to the statistical analysis of causal treatment effects, considering potential (not necessarily observed) outcomes. It is based on two critical assumptions (Rosenbaum & Rubin, 1983) — *stable unit treatment value assumption* (SUTVA) and *strong ignorability*. SUTVA implies that the distribution of potential outcomes for each individual is independent of the potential outcomes of other individuals (Cox & Cox, 1958). Strong ignorability includes two key parts related to the PS (or treatment assignment mechanism), defined here as $Pr(T_i = 1|X_i)$ where $X_i$ denotes the vector of observed pretreatment confounders used in the PS model. First, that there are no unobserved confounders in the PS model. We discuss how to address this assumption in more detail in Step 6. The second part states that each individual has a positive probability to be assigned to the treatment group ($0 < Pr(T_i = 1|X_i) < 1$) (Rosenbaum & Rubin, 1983). In observational studies, where the group assignment is likely to be determined by the medical and personal characteristics of the individual, it is possible to be able to predict the group allocation perfectly based on the baseline characteristics (Bergstra et al., 2019) if certain groups of individuals in the study only ended up in one group versus another. For this reason, it is important to report some summary statistics of the baseline covariates (like mean, standard deviation, minimum, maximum) per group to check that the distributions of the covariate values of the groups overlap. Unfortunately, there is no formal way to test this overlap assumption in a given sample. Instead, we recommend some simple checks that can be done to assess obvious areas of the covariate distributions where there is a lack of overlap. For example, overlap can be checked by comparing the minimum and maximum of the same covariate in the two groups or by distribution plots such as those shown in *Figure* 1.

If there are ranges of covariate values that only occur in one group, one might have to exclude some individuals from the analysis, such that the two groups are adequately overlapped, or consider other estimands, like the ATO (Li et al., 2018; Mlcoch et al., 2019), which measures the causal treatment effect on the region where there are representatives of both groups. Lack of overlap suggests that there is no uncertainty in the decision to treat individuals with a particular characteristic since individuals in this region are always assigned to one treatment group. The exclusion of participants with covariate values that do not overlap with the other group tends to create a more clinically useful target population. In the case of binary or/and categorical covariates, one should make sure that at least one individual is representing each category/level per treatment group — unless a category has no representative in either group.

We use *Figure* 1 to illustrate how plotting the density functions of a single covariate for each of the treatment groups can be a helpful way to identify obvious areas of the covariate distributions lacking overlap. It is apparent from *Figure* 1 that the support of the control group (red area) extends beyond the support of the treatment group — this is the area to the right of the vertical line. In such a case, one should consider how to handle this issue, either by estimating the balancing weight only for the common area — in practice, this could mean to exclude control group individuals with large values for this covariate — or by estimating the treatment effect only for the region of overlap (ATO) (Li et al., 2018).

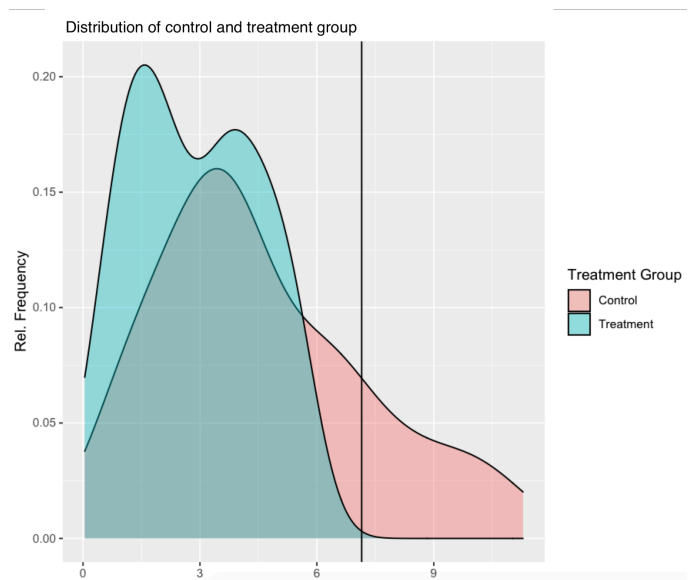

*Figure 1. Density plot of control and treatment group, where there is lack of overlap in the support of the two groups. The data used for this plot are artificial to demonstrate the issue.*

Consideration of overlap prior to estimation of the balancing and PS weights is an important part of the process of estimating causal treatment effects. Often researchers dive into estimation of the balancing and PS weights without careful consideration of overlap and find themselves frustrated by obtaining balancing weights that do not successfully balance their groups. If lack of overlap exists, it can be difficult to obtain high quality balancing and PS weights. Our Shiny application allows the user to examine the summary statistics of every confounder, as well as their density plots. There is an option to trim outliers for either upper or lower tail, or both, for one confounder at a time or more. This procedure enables the user to view the impact of removal of observations from one confounder on the remaining confounders when the observation is removed from the entire data set. It is quite common that observations with extremes in one confounder tend to be similarly extreme for other confounders.

3. **Step 3.** *Estimate the propensity score or balancing weights needed, ideally using multiple methods*

   There have been several articles comparing PS and balancing weighting methods (for example, (Abdia et al., 2017; Griffin et al., 2017; Mao et al., 2019; Setodji et al., 2017; Setoguchi et al., 2008)). Under different settings, different methods perform better, and this depends on the structure of the given data set — the sample size, the number of covariates to be balanced (especially relative to sample size), and the true underlying form of the treatment assignment model (e.g., linear versus non-parametric). If one has a small sample and the natural relation between the confounders and the allocation mechanism is a sigmoid function with main effects only, logistic regression seems a natural choice. In general, however, there is a lack of guidance on how to choose from the multitude of methods available in any particular analysis, in a way to achieve sufficient covariate balance without reducing the effective sample size more than necessary, to achieve (near) unbiased estimation of a causal treatment effect. It is not immediately obvious in any given setting which method is optimal. Thus, in this tutorial, we recommend the consideration of several methods and to make inference based on the one that achieves the best trade-off between balance and effective sample size (ESS) as explained in more detail in Step 4 (Ridgeway et al., 2017).

   *Assessment of PS Distribution Overlap.*
   There are several references that recommend checking the distribution of PS across the two groups (D'Agostino Jr, 1998; Elze et al., 2017; Li et al., 2018; McCaffrey et al., 2004) since lack of overlap could lead to extreme weights and potentially compromised estimation of the causal treatment effect. A region where the distributions of PS of the two groups do not overlap means that there is no uncertainty about the treatment allocation of individuals in this region — this is that the second assumption of the RCM does not hold. Regions of lack of overlap on PS distributions are regions in which the baseline characteristics of these individuals predict exactly the treatment allocation of an individual. This is why it is important to assess the overlap of the baseline covariates in step 2, to avoid such problematic behaviors in the

PS distributions. Given we address this issue in step 2, we consider that checking the PS distributions is not always necessary, as there should likely be no lack of overlap.

4. **Step 4.** *Assess balance and effective sample size for all methods and choose the best one for outcome analysis*

   Once PS or balancing weights are estimated, the balance (or comparability) among the groups needs to be assessed. The theory behind PS and balancing weights suggests that balance should be obtained on the full multivariate distribution of the observed confounders after one applies the weights. However, in practice, this is often not checked and it is challenging to properly test if the full multivariate distribution of the observed confounders balances between the treatment conditions.

   Here, we propose to use both the *standardized mean difference* (SMD) and the *Kolmogorov-Smirnov statistic* (KS) as a way to assess how comparable the two treatment groups are. The SMD allows one to assess the comparability of the means for each observed confounder while the KS statistic allows us to assess balance also in the tails of the distributions for a given confounder between the two treatment groups. These metrics are commonly used in the literature (Austin, 2009; Franklin et al., 2014; Gail & Green, 1976; Griffin et al., 2017; Setodji et al., 2017; Setoguchi et al., 2008; Zhao & others, 2019). Both metrics are explained in detail in Section 4.

   Additionally, we also carefully consider the impact of the weighting on the power of a study. The PS and balancing weights act in the same way as survey or sampling weights and add increased variability into the statistical models and treatment effect estimates. We can assess the impact of different PS and balancing weight methods by computing the ESS, which denotes the remaining sample size, after the reduction due to the variability in the weights. When balance across multiple methods is similar, one would naturally prefer to select as optimal the PS or balancing weights with the lowest reduction to the ESS. The ESS is also explained in more detail in Section 4.

5. **Step 5.** *Model the outcome and estimate the causal treatment effect.*

   Before proceeding with any outcome analyses, researchers must assess whether they will be able to estimate robust causal effects with their sample. To do so, they need to ensure adequate balance has been obtained with the PS or balancing weights being used in a given analysis. If the weights do not balance the groups being compared well SMD and KS statistic are both below 0.1, a study will not be able to robustly estimate the causal effect of the treatment of interest. It is important that researchers understand when this is not happening, the study can only be used to examine associations, rather than causal effects, since the findings will be less robust and must be caveated as such.

   Assuming one does have adequate balance and the largest possible ESS, there are several possible options for the estimation of the needed treatment effects. The simplest estimate is to compare weighted means between the treatment and control groups. Given the balancing weights $w_i$, for every individual $i$, an estimation of the causal treatment effect could be obtained by the formula:

$$AT\_ = \frac{\sum_{j \in C_1} w_i Y_i^{obs}}{\sum_{i \in C_1} w_i} - \frac{\sum_{j \in C_0} w_i Y_i^{obs}}{\sum_{i \in C_0} w_i}$$

where $C_1$ is the set of individuals in the treatment group, and $C_0$ the set of individuals in the control group.

However, it is more common practice to combine the weights with a multivariable regression adjustment that ideally includes all of the observed confounders used in the estimation of the weights (Austin, 2011; Ridgeway et al., 2017). That is, the PS or balancing weights are used as weights in an augmented regression model that also includes all observed confounders. For PS weights, this approach yields a *doubly robust estimator* of the causal treatment effect (Bang & Robins, 2005; Chattopadhyay et al., 2020; Kang et al., 2007; Zhao & Percival, 2016). The estimated treatment effect is consistent so long as one part of the doubly robust model is correct (i.e., either the PS weight model or the multivariable outcome model).Our Shiny application uses this approach to estimate the causal treatment effect of interest. For balancing weights, this doubly robustness property has not yet been established. However, the use of covariate adjustment in the regression model is still seen as useful for minimizing bias in the estimated treatment effect and increasing precision in the model. In cases where sample sizes are restricted and might not support fully adjusting for all covariates, it can be useful to control for a subset of the observed confounders, namely those that have the greatest lingering imbalance in the SMD or KS statistic.

6. **Step 6.** *Assess sensitivity of the results to unobserved confounding.*

   A key assumption in all weighted analyses is that we have not left out any potential unobserved confounders when estimating the PS or balancing weights. Unfortunately, as with the overlap assumption, the assumption of no unobserved confounders is impossible to test formally in practice. Yet, it is important to assess how robust a study's findings might be to unmeasured factors that have not been included in the weights. It is most common to utilize sensitivity analyses that assess the sensitivity of the estimated treatment effects and/or statistical significance of analysis to potential unobserved confounders. Despite their importance, such analyses are underutilized in the literature. Here, we showcase the use of a graphical tool to describe how sensitive both treatment effect estimates and statistical significance (as measured by the p-value) will be to an unobserved covariate (Griffin et al., 2020).

## Propensity Scoring and Balancing Weight Analysis Methods

We will now introduce some notation, which will be useful in the description of PS and balancing weight estimation algorithms. Consider a simple random sample of $N$ observations from a population $P$. For each unit $i$, we observe a binary treatment variable $T_i$ and a $K-$dimensional column vector of observed pre-treatment covariates $\mathbf{X}_i$. The PS is defined as the conditional probability of receiving the treatment given the covariates $\mathbf{X}_i$, i.e. $Pr(\mathrm{X}_i) = Pr(T_i = 1|\mathrm{X}_i)$. From (Rosenbaum & Rubin, 1983), the ignorability of treatment assignment says that the treatment assignment is ignorable given the (true) propensity score $Pr(X_i)$. This implies that the unbiased estimation of treatment effect is possible by conditioning on the PS alone instead of the entire covariate vector $\mathbf{X}_i$. However, in observational studies the PS must be estimated from the data set. Normally, one assumes a parametric PS model $Pr_\beta(X_i)$. That is, $Pr(T_i = 1|X_i) = Pr_\beta(X_i)$, where $\beta \in \delta$ is an $L$-dimensional column vector of unknown parameters.

In the following paragraphs, four commonly used algorithms for obtaining PS and balancing weights are described.

1. ***Logistic Regression***

    The simplest and most commonly used parametric model method to estimate the PS of each individual is logistic regression (LR) (Agresti, 2018) since treatment assignments are often binary. The basic logistic regression model for estimating the PS assumes that the *logit* of the probability of receiving treatment is equated with a linear combination of covariates $X_i^T\beta$. The coefficient vector $\beta$ is usually estimated with maximum likelihood estimation. The PS, in this case, is then computed from the estimated parameters as follows:

    $$Pr_\beta(X_i) = \frac{exp(X_i^T\beta)}{1 + exp(X_i^T\beta)}$$

    An extension of LR is multinomial logistic regression which could be used to estimate generalized PS if there are more than two treatment conditions. The main problem with the LR approach is that the PS model can easily be mis-specified, leading to biased estimates of treatment effects (Lee & Little, 2017; Pirracchio et al., 2015; Setoguchi et al., 2008; Wyss et al., 2014). It is nearly impossible to know the best way to specify the right hand side of the LR model.

2. ***Covariate Balance Propensity Score***

    To overcome the shortcoming caused by mis-specification of the model and create a parametric option focused on achieving good balance between the treatment groups, the covariate balancing PS method was developed by (Imai & Ratkovic, 2014).

    The authors used the covariate balancing property of parametric models by employing inverse PS weighting:

    $$E(\frac{T_i f(X_i)}{Pr_\beta(X_i)} - \frac{(1-T_i)f(X_i)}{1 - Pr_\beta(X_i)}) = 0$$

    where $f(X_i)$ is an $M$-dimensional vector-values measurable function of $X_i$ specified by the researcher. For instance, if $f(X_i)$ is the first derivative of $\pi_\beta(X_i)$, the assumed parametric model is logistic.

    This method has the advantage of being robust to mild model mis-specification with regard to balancing confounders compared to direct maximum likelihood estimation used in a standard LR. Additionally, the CBPS method can improve the covariate

balance in observed data sets and improve the accuracy of estimated treatment effects over parametric models even if there is no mis-specification (Choi & O'Reilly, 2019; Imai & Ratkovic, 2014; Setodji et al., 2017; Wyss et al., 2014; Xie et al., 2019).

The covariate balance method uses a generalized method of moments or an empirical likelihood estimation approach to find estimates that come closest to optimizing the likelihood function while concurrently meeting the balance condition for the weighted means of the covariates in the parameter estimation procedure. Even though the default set of restrictions (expressed through $f(\cdot)$) targets to balance the first moment (this is to minimize the SMD (Austin & Stuart, 2015) — see section *Measures to Evaluate Balance*), one could apply higher moments restrictions. A second-order restriction, or higher, could push the *KS statistic* (Gail & Green, 1976) towards 0. One could do so by transforming each continuous confounder via an orthogonal polynomial of degree 2 (Huang MY Vegetabile B, n.d.). This procedure increases the number of baseline covariates included in the treatment allocation model (times $m$, where $m$ is the order of the orthogonal polynomial), thus special caution should be taken here since this could make the achievement of adequate balance harder.

3. ***Generalized Boosted Model***

    GBM is a flexible, nonparametric machine learning approach to estimating PS weights. It predicts the binary treatment indicator by fitting a piecewise-constant model, constructed as a combination of simple regression trees (Burgette, McCaffrey, and Griffin in press, (Burgette et al., 2015; Ridgeway, 1999)), namely *Recursive Partitioning Algorithms* and *Boosting*. To develop the PS model, GBM uses an iterative, *forward stagewise additive algorithm*. Starting with the PS equal to the average of treatment assignment on the sample, such an algorithm starts by fitting a simple regression tree to the data to predict treatment from the covariates by maximizing the following function.

    $$l(x) = \sum_{i=1}^{N} T_i \, g(X_i) - log(1 + exp(g(X_i))),$$

    where $g(X_i)$ is the $logit$ of treatment assignment. Then, at each additional step of the algorithm, a new simple regression tree is added to the model from the previous iterations without changing any of the previous regression tree fits. The new tree is chosen to provide the best fit to the residuals of the model from the previous iteration. This chosen tree also provides the greatest increase to the log-likelihood for the data. When combining trees, the predictions from each tree are shrunken by a scalar less than one to improve the smoothness of the resulting piecewise-constant model and the overall fit.

    The number of iterations that are performed by the algorithm or the number of trees in the model determines the model's complexity. When choosing the number of iterations to yield the final PS model, one must pick a value that balances between underfitting (i.e. not capturing important features of the data) and overfitting the data. One selects the *final* model of the treatment indicator (and correspondingly, the PS

and PS weights needed for analysis) by selecting a particular number of iterations considered *optimal* where optimization is done based on achieving the best balance — the most commonly used editions of the algorithms are the ones who target to optimize the mean standardized mean difference ($GMB_{ES}$) and the maximum Kolmogorov-Smirnov statistic ($GBM_{KS}$).

A detailed tutorial of GBM for PS estimation, balance evaluation, and treatment effect estimation in R software is discussed in (Ridgeway et al., 2017). GBM methods could also be used when there are more than two treatments (McCaffrey et al., 2013).

4. **Entropy Balancing**

   *Entropy balancing* (EB) is a method that aims to estimate the weights directly rather than via the PS of the individuals. The method promises to achieve exact balance on as many moments as defined by the user.

   Assuming that one is interested in estimating the ATT, the quantity that is tricky to estimate is $E[Y_{i|T_0}|T_i = 1]$ (this is the expected outcome value of individuals in the control groups, considered they received the treatment), since these values are not observed. Thus, an estimation of the above quantity is

   $$E[\widehat{Y_0|T} = 1] = \frac{\sum_{C_0} w_i^0 Y_i^{obs}}{\sum_{i=1}^{n_0} w_i^0},$$

   where $w_i$ are the balance weights and need to be estimated, and $n_0$ is the number of inviduals in the control group. The entropy balancing method calculates weights through a reweighting scheme (until adequate balance in the pre-advised moments is achieved), while at the same time attempting to satisfy a set of *balance and normalizing constraints* — equations (3), (4) and (5) in (Hainmueller, 2012). This set of restrictions corresponds to matching the first $k$ moments of the distributions of the two groups ($k$ is defined by the user). Entropy balancing re-weighting schemes can be considered as generalizations of the traditionally used Inverse Probability Weighting (Hainmueller, 2012; Zhao & others, 2019; Zhao & Percival, 2016). However, the weights here are not defined by the PS, thus a direct estimation of PS is not feasible.

We note that the PS weights $w_i$ are defined in a different way, depending on the components of equations (1), (2) or (3) we wish to estimate. The weights are defined as follows:

$$w_i = T_i \frac{1}{\Pr(x_i)} + (1 - T_i) \frac{1}{1 - \Pr(x_i)}, for\ ATE, (6)$$

$$w_i = T_i + (1 - T_i) \frac{\Pr(x_i)}{1 - \Pr(x_i)}, for\ ATT, (7)$$

$$w_i = T_i \frac{1 - \Pr(x_i)}{\Pr(x_i)} + (1 - T_i), for\ ATC. (8)$$

*Logistic regression*, *GBM* and *CBPS* produce PS, which we transform into balancing weights using Equations (6), (7) and (8), whereas *entropy balancing* estimates the weights directly.

## Measures to Evaluate Balance

Once the estimation of weights is available, it is important to evaluate the balance on the two groups achieved. To do so we will utilize three key metrics: *standardized mean difference* (SMD), *Kolmogorov-Smirnov statistic* (KS) and *effective sample size* (ESS). The first two are measures of the balance of the two distributions, while the latter is a measure of the sample power lost by weighting.

### Standardized Mean Difference

SMD (Austin, 2009; Austin & Stuart, 2015; Franklin et al., 2014) is a measure of the distance of the means of two groups. It is defined as the difference of the means, divided by an estimate of the standard deviation, for a given covariate — depending on the causal treatment effect one wishes to estimate. For ATE, it is formally defined as:

$$SMD = \frac{\bar{X}_{treatment}^{weighted} - \bar{X}_{control}^{weighted}}{\sqrt{\frac{sd_{treatment,weighted}^2 + sd_{control,weighted}^2}{2}}},$$

where $X_{treatment}^{weighted}$ and $X_{control}^{weighted}$ are the weighted sample means. A weighted mean is defined as $\bar{x}^{weighted} = \frac{\sum_i w_i x_i}{\sum_i w_i}$, where $w_i$ is the weight of observation $i$ — in our context this is the PS and balancing weights. $sd_{treatment,weighted}$ and $sd_{control,weighted}$ are the weighted standard deviations of treatment and control group. A weighted standard deviation is defined as $sd_{weighted}^2 = \frac{\sum_i w_i}{(\sum_i w_i)^2 - \sum_i w_i^2} \sum_i w_i (x_i - \bar{x}^{weighted})^2$.

The lower the SMD value, the better the balance achieved. In the recent literature, 0.1 is recommended as a threshold (Austin, 2009; Austin et al., 2007; Austin & Mamdani, 2006; Griffin et al., 2014, 2017; Ho et al., 2007; Stuart et al., 2013; Zhao & others, 2019) to define groups as balanced, while in the past 0.2 was considered as adequate balance. (Abdia et al., 2017), however, provides strong evidence that such a threshold is very liberal, and does not guarantee an unbiased estimator of the causal treatment effect. By definition, SMD quantifies the similarity of the means of the two groups for each variable, thus it expresses how close the two mean values are.

### Kolmogorov-Smirnov Statistic

KS is a statistical test (Gail & Green, 1976) which checks the hypothesis that the two samples are from the same distribution. Its test statistic is formally defined as

$$KS = \max_z \left| F_{emp}^{treatment}(z) - F_{emp}^{control}(z) \right|, (10)$$

where $F_{emp}^{treatment}(\cdot)$ and $F_{emp}^{control}(\cdot)$ are the empirical distributions of treatment and control, respectively. The empirical distribution of a sample $x_1, x_2, \ldots, x_n$, is:

$$F_{emp}(x) = \frac{\#x_i \leq x}{n}. (11)$$

By definition, the *KS statistic* takes values in [0,1], and the lower the value, the closer the two distributions. Unlike SMD, it is a measure that quantifies the similarity of the entire distribution of the two groups, rather than the means only. There is no clear guidance on what is the best threshold for the KS but values over 0.1 would be consider notable large and thus, we propose to use 0.1 as the threshold for balance for the KS as well as the SMD.

The definitions in equations (10) and (11) are not considering weights, but this could be easily incorporated, if we replace the sample $\{x_i\}_i$ with $\{z_i\}_i$, which is the weighted version of $\{x_i\}_i$ — $z_i = \frac{w_i}{\sum_i w_i} x_i$, for every $i$.

*Effective Sample Size*

Given the weights of each group, the ESS (Ridgeway et al., 2017; Shook-Sa & Hudgens, 2020) is defined as:

$$ESS = \frac{(\sum_{i \in C} w_i)^2}{\sum_{i \in C} w_i^2}, (12)$$

where $w_i$ are the weights of the group $C$ — this could be either treatment or control group, if we are interested in the estimation of ATC or ATT, respectively, or the entire sample, if we wish to estimate ATE.

ESS expresses the number of observations from a random sample one would have to use to obtain an estimate with the same variance as the one obtained from the weighted group. It, therefore, can be used to help understand the power/precision a study has after using PS or balancing weights. The ESS will always be smaller than the *original* sample size.

## Study Data

As noted, the motivating case study data used in this tutorial comes from longitudinal observational study data on adolescents receiving substance use disorder (SUD) treatment, who were administered the Global Appraisal of Individual Needs (GAIN) biopsychosocial assessment instrument (Dennis et al., 2003) on a recurring basis. The GAIN was routinely collected by 178 adolescent SUD treatment sites funded by the Center for Substance Abuse Treatment (CSAT) between 1997 and 2012, including the sites serving our A-CRA and MET/CBT5 groups. In total, we have 4968 adolescents from the A-CRA group and 5184 from the MET/CBT5 group with GAIN data at baseline and 6-months post-baseline.

Since the allocation mechanism between the two treatment groups is not random, we will control for 20 pretreatment confounders (for explanation of these covariates see Appendix

(Dennis et al., 2003)) in our analysis. In brief, they cover a range of baseline confounders including sociodemographic factors like age, race/ethnicity, and gender as well as baseline measures of substance use, mental health and environmental risk. These confounders were chosen based on prior research with experts in the field of SUD (Grant et al., 2020) as well as the literature as key potential predictors of outcomes in this sample (Griffin et al., 2012; Ramchand et al., 2011, 2014, 2015; Schuler et al., 2014). While dealing with variable selection for PS and outcome model is beyond the goal of this current tutorial , we note that it is important to ensure selection of all important potential confounders prior to performing the analyses outlined in this tutorial. Our key outcome is the total number of days abstinent in the 90-days prior to the 6-month follow-up.

*Table* 1 shows the baseline characteristics of youth in our two groups. As shown, the groups showed differences on 17 out of our 20 pretreatment confounders (i.e. they have SMD and/or KS > 0.1). In particular, *Table* 1 suggests A-CRA more likely to be non-white. They also had higher mean levels of substance use and mental health problems (including traumatic stress, emotional problems, internal mental distress and behavioral complexity). Youth in A-CRA were also more likely to have had past substance use and mental health treatment than youth in the MET/CBT5 group. Finally, youth in the A-CRA group were less likely to have parental involvement at home and more likely to have higher levels of risk in their living and social environments. In light of these differences between the treatment groups on the baseline pretreatment confounders, it is clear this case study will benefit from use of PS or balancing weights to remove the differences between the two groups on these observed baseline factors. We will now walk through the six needed steps of this tutorial to obtain balancing weights that ensure the groups are more comparable and showcase how to robustly examine the estimated treatment effects from the optimal set of weights.

We note that the original case study data analysis included 90 imputed datasets. We randomly selected one of the 90 imputed data sets to allow for easier illustration of the steps of this tutorial. When doing PS or balance weighting with imputed data, it is necessary to repeat steps 4 and 5 of the tutorial for each imputed data set and then aggregate results across imputations to get the final estimated effect of treatment.

|  | MET/CBT5 | A-CRA | Balance | |
|---|---|---|---|---|
|  | **Mean** | **Mean** | **SMD** | **KS** |
| **Age** | 15.42 | 15.62 | 0.15 | 0.07 |
| **Traumatic Stress Scale** | 1.79 | 2.26 | 0.14 | 0.06 |
| **Substance Frequency Scale** | 10.40 | 11.88 | 0.12 | 0.05 |
| **Depressive Symptom Scale** | 2.59 | 2.80 | 0.08 | 0.04 |
| **Emotional Problems Scale** | 20.73 | 25.48 | 0.25 | 0.12 |
| Internal Mental Distress Scale | 7.49 | 8.51 | 0.12 | 0.06 |
| Behavior Complexity Scale | 9.62 | 11.02 | 0.17 | 0.07 |
| Adjusted Days Abstinent (past 90 days) | 55.11 | 47.76 | 0.22 | 0.10 |
| In recovery | 0.25 | 0.25 | 0.00 | 0.00 |
| Mental Health Treatment (past 90 days) | 0.18 | 0.23 | 0.11 | 0.05 |
| Substance Abuse TX Index | 1.44 | 8.01 | 0.39 | 0.14 |
| General Conflict Tactic Scale | 2.85 | 3.23 | 0.14 | 0.07 |
| Continued Substance Use Despite Prior Tx | 0.09 | 0.11 | 0.07 | 0.10 |
| Recovery Environmental Risk Index | 47.85 | 41.94 | 0.35 | 0.16 |
| Parent Activity Index | 3.57 | 3.26 | 0.23 | 0.11 |
| Substance Use Dependence (past year) | 3.64 | 4.39 | 0.22 | 0.10 |
| Living Environmental Risk Index | 10.32 | 10.78 | 0.15 | 0.06 |
| Social Environmental Risk Index | 12.80 | 13.22 | 0.10 | 0.05 |
| | % per group | | Balance | |
| | MET/CBT5 | A-CRA | SMD | KS |
| Race:White | 50.8% | 31.9% | 0.41 | 0.19 |
| Race:Hispanic | 22.5% | 31.8% | 0.21 | 0.09 |
| Race:African American | 10.1% | 15.9% | 0.17 | 0.06 |
| Race:other | 16.6% | 20.5% | 0.10 | 0.04 |
| Gender:Female | 26.75% | 29.9% | 0.07 | 0.03 |

*Table 1.* Mean values of baseline characteristics per group.

## Obtaining PS Weighting Estimates and Assessing their Performance

The Covariate Balancing & Weighting Web App (*CoBWeb*) was developed by the main author, as part of the article, to help users apply the procedure described in *section The 6 Key Steps towards Estimating Causal Treatment Effects*. The app implements all the steps, in a user-friendly way. *CoBWeb* is freely available at https://andreasmarkoulidakis.shinyapps.io/cobweb/.

General guidance about the implementation of the *key steps* in the app is described in the Appendix.

### Step 1: Choose which estimand one is interested in (ATE, ATT, ATC)

This analysis will focus on estimating the ATE in order to allow us to generalize the findings for our sample of youth enrolled in both treatments.

### Step 2: Assess sample for any obvious overlap concerns and adjust sample as needed

Next, we carefully assess the overlap between the control (MET/CBT5) and treatment (A-CRA) groups. Notably, there were no concerns about overlap for any of the 20 pretreatment confounders used in our analysis. *Figure* 2 provides a view of the overlap density plots for three of the confounders – age, substance use at baseline, and the depression scale – to illustrate the patterns seen for all 20 confounders. As shown, there are no concerns about lack of overlap for these confounders. Both groups have individuals with a similar range of observed values for each confounder. As it is apparent, the treatment groups are well overlapped, thus there is no apparent need to trim the data set. We will continue the analysis without the removal of any outliers.

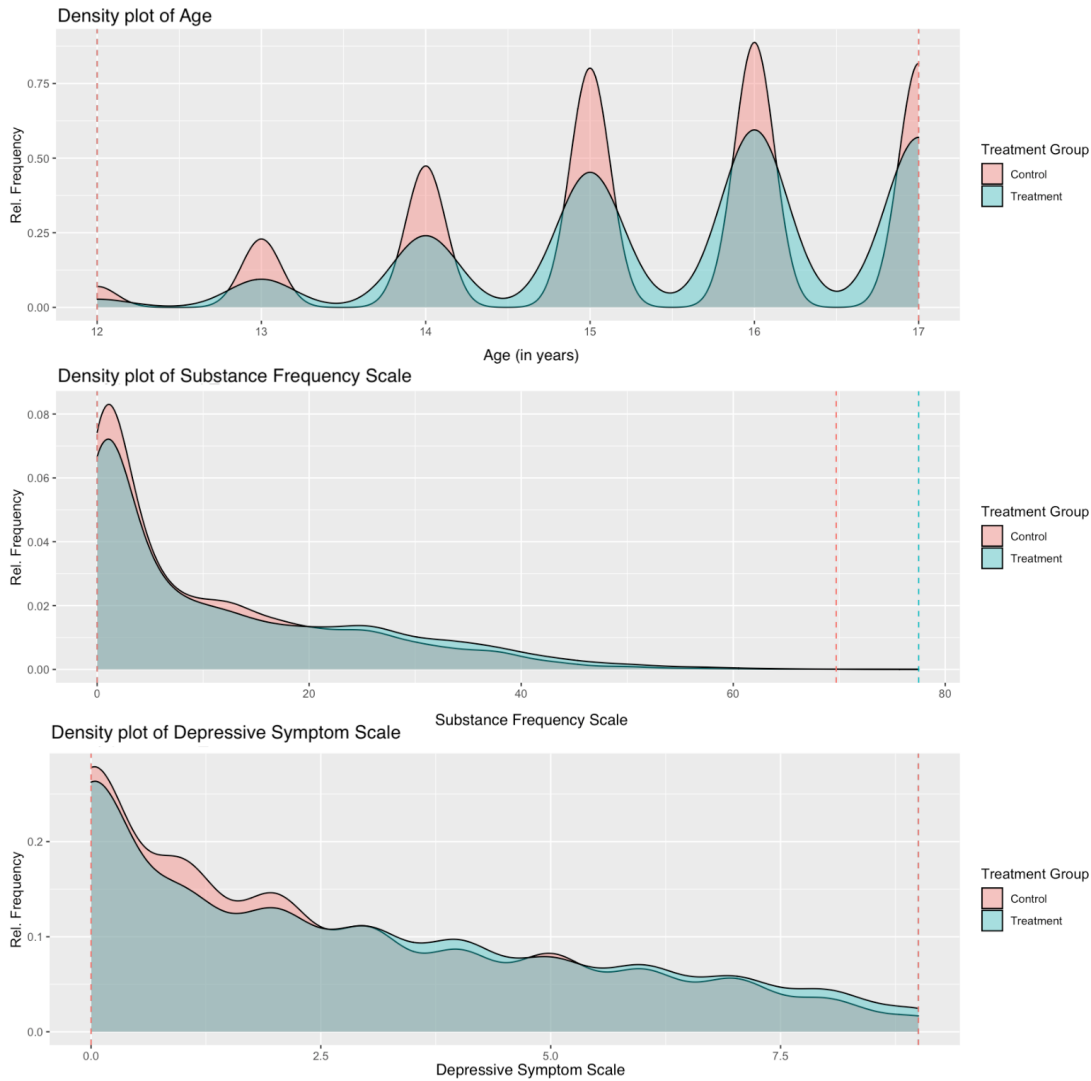

**Figure 2.** Density plots of *Age*, *Substance Frequency Scale* and *Depressive Symptom Scale* — light red is the control group (A-CRA) and with light blue the treatment group (MET/CBT5).

## Step 3: Estimation of propensity scores or balancing weights, ideally using multiple methods

Now we estimate the balancing and PS weights using the four methods presented in section *Propensity Scoring and Balancing Weight Analysis Methods* (this corresponds to 9 algorithms in total). Here, we are interested in the estimation of ATE, thus the relevant formula for estimation of needed PS weights from LR, GBM and CBPS is given by Equation (6). EB estimates the balancing weights directly and, thus, no transformation is done.

## Step 4: Assess balance and effective sample size for all methods and choose the best one for outcome analysis

Having estimated the balancing weights, we next assess balance after applying the weights from each method.

*Table* 2 reports the mean and the maximum SMD and KS values per method, as well as the ESS per method.

| SMD | | | | | | | | | | |
|---|---|---|---|---|---|---|---|---|---|---|
| | Unweighted | LR | GBM | GBM | CBPS #1 | CBPS #2 | CBPS #3 | EB #1 | EB #2 | EB #3 |
| mean | 0.15 | 0.02 | 0.03 | 0.03 | 0.01 | 0.01 | 0.02 | 0.00 | 0.00 | 0.00 |
| max | 0.39 | 0.13 | 0.07 | 0.07 | 0.02 | 0.05 | 0.06 | 0.00 | 0.00 | 0.00 |
| KS Statistic | | | | | | | | | | |
| | Unweighted | LR | GBM | GBM | CBPS #1 | CBPS #2 | CBPS #3 | EB #1 | EB #2 | EB #3 |
| mean | 0.07 | 0.02 | 0.02 | 0.02 | 0.02 | 0.01 | 0.01 | 0.01 | 0.01 | 0.01 |
| max | 0.16 | 0.05 | 0.03 | 0.03 | 0.06 | 0.05 | 0.04 | 0.05 | 0.04 | 0.03 |
| ESS | | | | | | | | | | |
| | Unweighted | LR | GBM | GBM | CBPS #1 | CBPS #2 | CBPS #3 | EB #1 | EB #2 | EB #3 |
| ESS | 10152 | 6062 | 7752 | 7877 | 7534 | 7564 | 7711 | 8071 | 7838 | 7743 |
| % of sample | 100% | 60% | 76% | 78% | 74% | 75% | 76% | 80% | 77% | 76% |

**Table 2.** Standardized Mean Difference (SMD), Kolmogorov-Smirnov Statistic (KS) and Effective Sample Size (ESS) per estimation method. For SMD and KS 0.1 is considered as a widely accepted threshold for good balance, while ESS should be as close to unweighted as possible.

As observed in *Table* 2, *EB* achieves the best balance in terms of SMD (absolutely 0 for all covariates — since maximum SMD is equal to 0), followed closely by all other algorithms. Only the logistic regression model has a maximum SMD (0.13) above the widely accepted threshold of 0.1 (Ridgeway et al., 2017), which suggests there is lingering meaningful imbalance for the PS weights associated with this method.

In terms of KS statistic, the lowest values are reported by *GBM* algorithms — maximum KS value equal to 0.03 for both algorithms. All algorithms though, achieve adequate balance when using the 0.1 threshold for balance.

Taken all together, the majority of our approaches achieved successful balance after weighting except for logistic regression. All have maximum SMD and KS below the 0.1 cut-point — but LR.

At this point, it would be advisable to choose the optimal algorithm as the one with the best combination of lowest maximum KS and the largest ESS — the sample size retained after the weighting process. Although, with the exception of LR, differences between the algorithms are small, $GBM_{KS}$ appears optimal by this criterion. Thus, balancing weights estimated by this algorithm will be used for the outcome analysis.

## Step 5: Model outcome and estimate the causal treatment effect

Once adequate balance is achieved, the next step is to estimate the causal treatment effect, using the balancing weights identified as optimal in the previous step.

To do so, we fit a weighted linear regression on *Adjusted Days Abstinent (past 90 days)* as the outcome, measured at 6-months after intake, considering as predictors all the other pretreatment covariates shown in Table 1 (including *Adjusted Days Abstinent (past 90 days)* measured at baseline), and the treatment status (A-CRA). The coefficient of treatment status represents the estimand of interest. The treatment effect is estimated to be equal to 1.23, with a 95% confidence interval equal to (0.00,2.57) and an associated p-value of 0.0498 ($\simeq 0.05$), which means that the effect of the treatment is considered to be small and marginally statistically significant (at 5% level of significance), suggesting that youth in A-CRA had slightly higher mean days abstinent at 6-months post intake (here approximately equal to 1.2 days more with a 95% confidence interval equal to (0,2.6) days.

## Step 6: Assess sensitivity of the results to unobserved confounding

In the final step of the tutorial, we carefully consider how sensitive findings from the case study might be to potential omitted pretreatment confounders. Specifically, we use a graphical tool (Pane et al., n.d.) which describes how sensitive both treatment effect estimates and statistical significance (as measured by the p-value) will be to an unobserved covariate.

The output is a contour plot (*Figure* 3), which shows how both the estimated treatment effect (solid contours) and p-value (dashed contours) would change as a function of an unobserved confounder whose association with the treatment indicator is expressed through an effect size or SMD (x-axis) and whose relationship with the outcome is expressed as a correlation (y-axis). More specifically, the x-axis displays the *Association with Treatment Indicator* — this is the unweighted standardized mean difference of the potential unobserved confounder between the treatment and control groups —, and the y-axis displays the *Absolute Association with the Outcome Covariate* — this is the absolute correlation of the unobserved confounder with the outcome covariate. The plot also provides the estimated treatment effect and the associated p-value (caption) from the original analysis, which is an indicator of the significance of the effect.

The solid black lines of the contour plot show how the size of the treatment effect changes as a function of the unobserved confounder's relationship with treatment assignment and the outcome, here showing that the estimated treatment effect will get larger as we move to the left from 0 on the x-axis but will decrease to 0 and then move negative as we move right from 0 on the x-axis.

The red dashed contours show how the size of the p-value changes as a function of the unobserved confounder focusing on commonly used cut-offs. Since our finding had a p-value right at the 0.05 threshold, we find out that the statistical significance only gets stronger as we move left from 0 on the x-axis and the estimated treatment effect gets larger. As we move right from 0 on the x-axis, we quickly move into an area of statistical non-significance before again crossing back into an area where the p-value will be less than 0.05 as our treatment effect starts getting larger in the negative direction.

The blue dots help to put the findings into perspective by plotting the observed correlations between our observed pretreatment confounders and the outcome and treatment indicator. Here, these all fall into the range where our treatment effect is close to 0 and has a non-significant p-value. This suggests it is highly possible we left out an important unobserved confounder that might quickly make our estimated treatment effect basically null. An example unobserved confounder that might have correlations similar to the observed confounders shown would be family history of substance use or mental health problems.

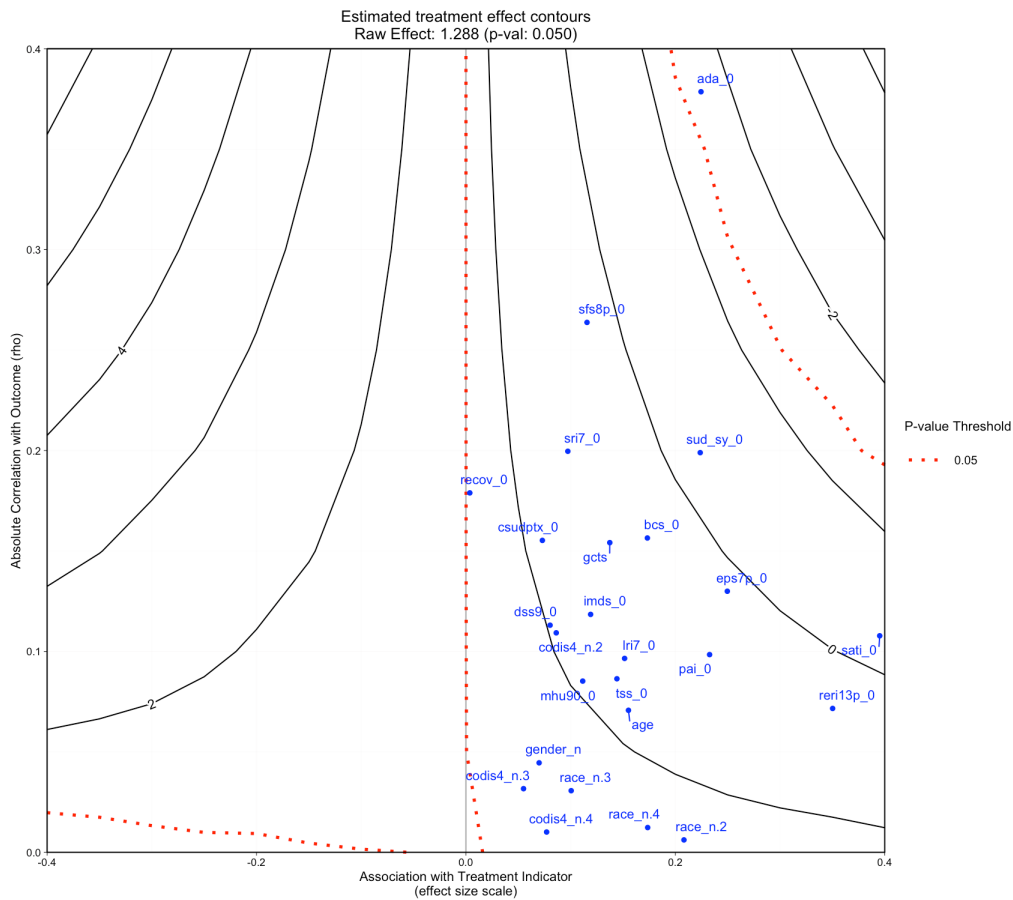

**Figure 3.** The x-axis indicates the SMD of potential confounders, and the y-axis the correlation with the outcome. The *black solid* contours represent the size of the treatment effect, while the red dashed ones represent the p-value cut-offs (0.05 level). The blue dots represent the association of the confounders with the treatment and the outcome.

Our findings in this case study are clearly *very sensitive* to unobserved confounders. As a consequence, considering an unobserved confounder would be negatively associated to the treatment (negative SMD on the baseline) on the range [0,0.4] — possessed on the left side of the graph—, and a low/moderate association to the outcome (correlation with the outcome in the range [0,0.4]), would influence the estimated causal treatment effect towards 0 value, with lower significance, and thus provide evidence that the days abstinent under A-CRA would not be considered significantly different compared to MET/CBT5.

## Discussion/Conclusion

In this paper, we present a step-by-step guide to making inference on causal treatment effects using observational data. This tutorial represents an advancement over prior work (Ali et al., 2016; Garrido et al., 2014; Lee & Little, 2017; Olmos & Govindasamy, 2015) in that we explicitly deal with addressing two key assumptions for PS methods (overlap and unobserved confounding) and recommend the use of multiple estimators for the potential PS and balancing weights in order to ultimately select the best method for a given study. It is difficult to be able to project which PS or balancing method will do best in any given study and thus, we believe it is better for it to become standard practice (before looking at any outcome models) to use multiple methods and carefully compare balance and ESS to select the one that is optimal.

Observational studies are widely used in the research of causal treatment effects. Thus, the accurate estimation of the effect of interest is of primal importance. Since the treatment and control groups are not balanced a priori, unlike in a (large) RCT, PS and balancing weights are a useful tool for every researcher who wishes to make inference based on observational data.

Observational studies, though, can pose several challenges, including extreme values on baseline covariates in one group, as well as limitations related to sample size. When dealing with small samples, a case which is quite often when studying rare diseases, one should be careful with the number of confounders which one attempts to balance, because there is always the danger of model overfitting. Variable selection is beyond the scope of this tutorial. We used expert input and past literature to choose the covariates we are interested to control for in our case study, however, there is some literature covering the topic of variable selection for PS models (Brookhart et al., 2006; Hirano & Imbens, 2001).

It is crucial to investigate the data *a priori* for potential overlapping or outliers issues, and adjust as needed. We highly recommend considering more than one method when it comes to algorithms that produce balancing weights and then evaluate the balance for each method and select the best performing one. Following outcome analysis it is important to perform an analysis of the sensitivity of the outcome estimation and significance to unobserved confounding, to access the generalization abilities of the outcome results.

## Funding Support

DOMINO-HD (which funds the Ph.D. of the main author) is funded through the EU Joint Program for Neurodegenerative Disease Research with UK funding from Alzheimer's Society and Jacques and Gloria Gossweiler Foundation. The Centre for Trials Research, Cardiff University receives infrastructure funding from Health and Care Research Wales. This work was also supported by Medical Research Council (UK) grant MR/L010305/1. Funding was also provided by grant R01DA045049 (PI Griffin) through the National Institute of Drug Abuse.

## Acknowledgments

The development of this manuscript was also supported by the Center for Substance Abuse Treatment (CSAT), Substance Abuse and Mental Health Services Administration (SAMHA) [#270-2003-00006, #270-2003-00006, and #270-2007-00004C] using data provided by the following grantees: TI-15413, TI-15415, TI-15421, TI-15433, TI-15438, TI-15446, TI-15447, TI-15458, TI-15461, TI-15466, TI-15467, TI-15469, TI-15475, TI-15478, TI-15479, TI-15481, TI-15483 TI-15485, TI-15486, TI-15489, TI-15511, TI-15514, TI-15524, TI-15527, TI-15545, TI-15562 TI-15577 TI-15584, TI-15586, TI-15670, TI-15671, TI-15672, TI-15674, TI-15677, TI-15678, TI-15682, TI-15686, TI-17589, TI-17604, TI-17605, TI-17638, TI-17646, TI-17648, TI-17673, TI-17702, TI-17719, TI-17728, TI-17742, TI-17744, TI-17751, TI-17755, TI-17761, TI-17763, TI-17765, TI-17769, TI-17775, TI-17779, TI-17786, TI-17788, TI-17812, TI-17817, TI-17821, TI-17825, TI-17830, TI-17831, TI-17847, TI-17864, TI-20759, TI-20781, TI-20798, TI-20806, TI-20827, TI-20828, TI-20847, TI-20852, TI-20865, TI-20870, TI-20910, TI-20946, TI-23174, TI-23186, TI-23188, TI- 23195, TI-23196, TI-23197, TI-23200, TI-23202, TI-23204, TI-23206, TI-23224, TI-23244, TI-23247, TI-23265, TI-23270, TI-23276, TI-23278, TI-23279, TI-23296, TI-23298, TI-23304, TI-23310, TI-23311, TI-23312, TI-23316, TI-23322 ,TI-23323 ,TI-23325 ,TI-23336 ,TI-23345 ,TI-23346, TI-23348. The authors thank these agencies, grantees, and their participants for agreeing to share their data to support this secondary analysis. The opinions about this data are those of the authors and do not reflect official positions of the government or individual agencies. Please direct correspondence about the data to Beth Ann Griffin, 1200 South Hayes Street, Arlington, VA 22202, USA bethg@rand.org, (703) 413-1100x5188.

# CoBWeb

*CoBWeb* is freely available at https://andreasmarkoulidakis.shinyapps.io/cobweb/

Initially, the user has to upload a data set using the *Data* panel. A summary of each covariate will appear (mean, standard deviation, median, minimum and maximum value), as well as the first few rows of the raw data.

### Choose which estimand one is interested in (ATE, ATT, ATC).

Moving to the *Model Set Up* panel, there will appear some multiple-choice questions to define the *Treatment Status* covariate, any other binary, continuous and categorical confounders, the outcome covariate, — all the covariates included in the dataset will be available options — as well as a choice of the estimand that the user wishes to estimate (ATE, ATT, ATC).

### Assess sample for any obvious overlap concerns and adjust sample as needed

The *Dealing with Outliers* panel provides a view of the per-group summary statistics of each covariate — mean, sd, median, min, and max for three of the confounders. Under the sub-panel, the user will have a glance at the differences of the distributions of the two treatment groups, for each covariate, which provides a graphic way to access any obvious overlapping concerns — options for removing outliers are available here.

### Estimation of propensity scores or balancing weights, ideally using multiple methods

Next, we estimate balancing weights using the four methods presented in Section *Propensity Score and Balancing Weight Analysis Methods* (9 algorithms in total).This is automatically done in the app.

### Assess balance and effective sample size for all methods and choose the best one for outcome analysis

In the next panel of the app, called *Balance Evaluation*, the SMD, KS statistic, and ESS (raw and as a percentage of the original size) are reported for each PS and balancing weights algorithm, alongside some recommendations about which algorithm performs better in each measure. At this stage, the user can choose which algorithm they wish to proceed with for the outcome analysis.

### Model outcome and estimate the causal treatment effect

The *Outcome Analysis* panel is split into two sub-panels — *Treatment Effect Estimation* and *Sensitivity Analysis*. The former provides a table with the estimation of the causal treatment estimation effect and its statistical significance.

### Assess sensitivity of the results to unobserved confounding

If one wishes to proceed with the sensitivity analysis (rather than stopping after step 5) — which is something we highly recommend — this can be done in the *Sensitivity Analysis* sub-panel of the *Outcome Analysis* panel. Once this is done, the user can download the sensitivity analysis graph, and move to the final panel (*Before you go!*).

## Before you go!

In the final panel (*Before you go!*) one can download an enhanced report (including the sensitivity analysis findings), and the final data with the weights.

# Density Plots

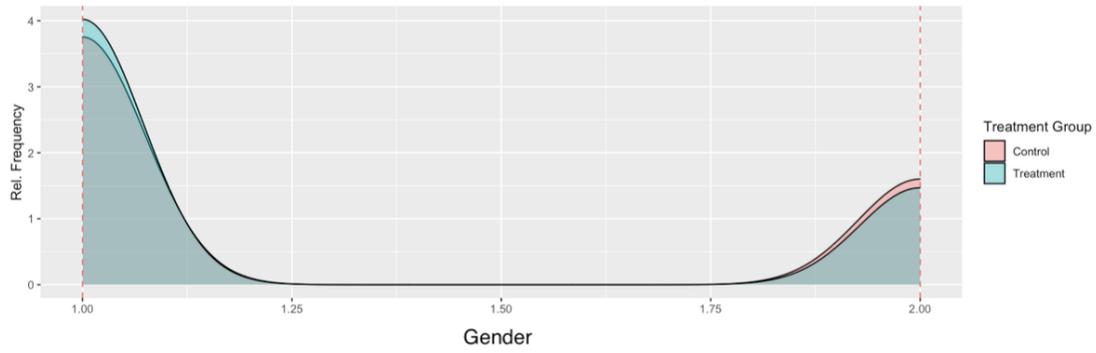

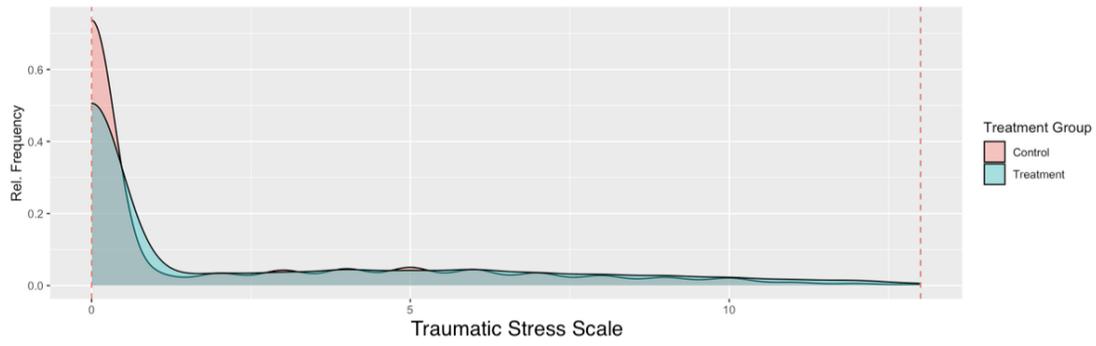

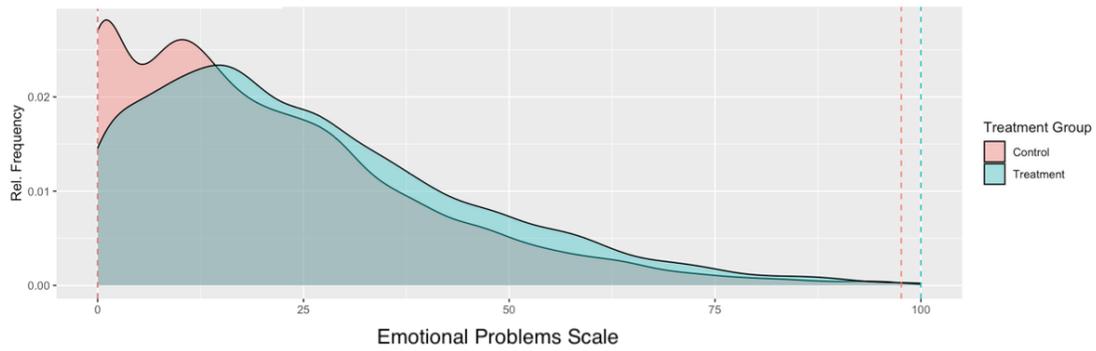

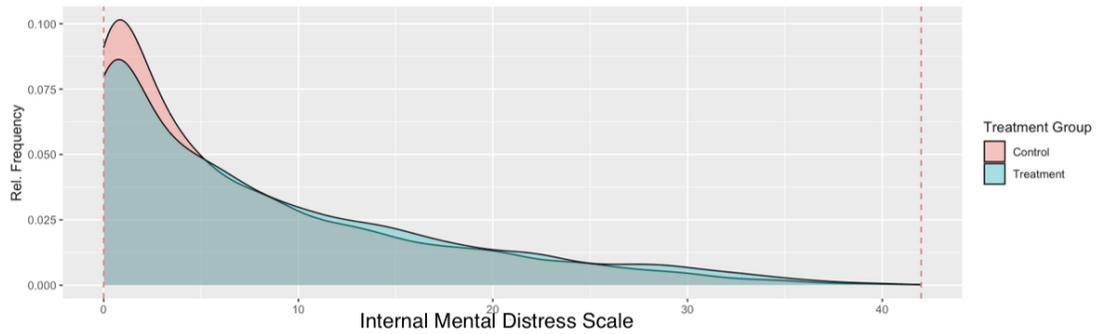

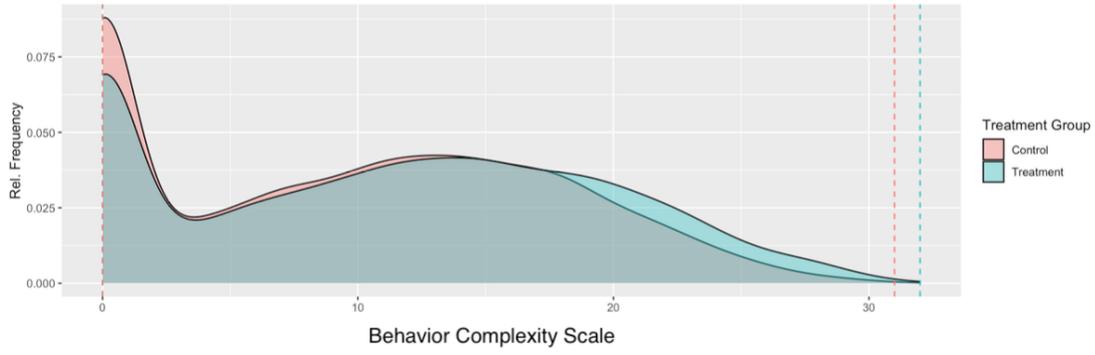
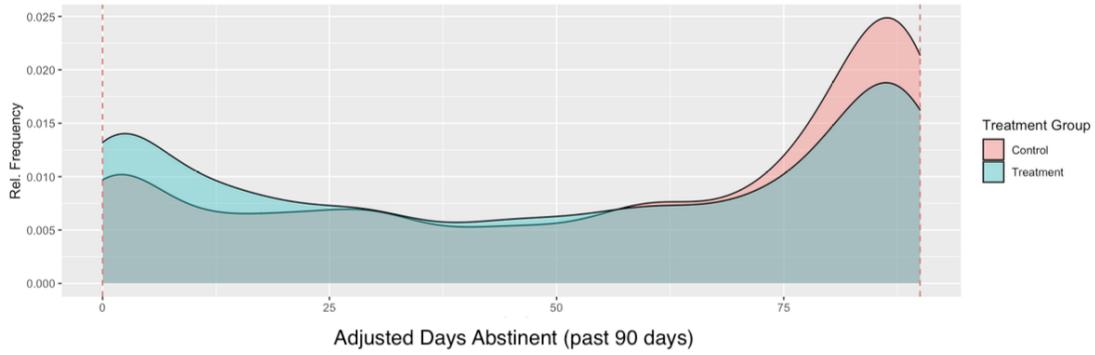
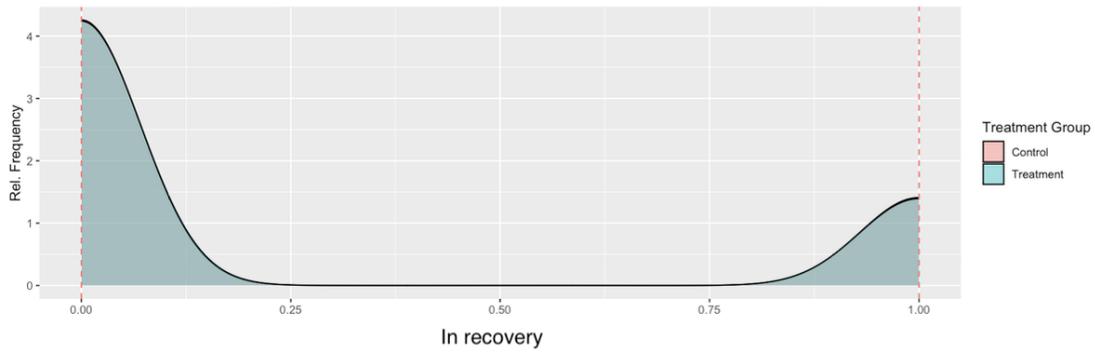
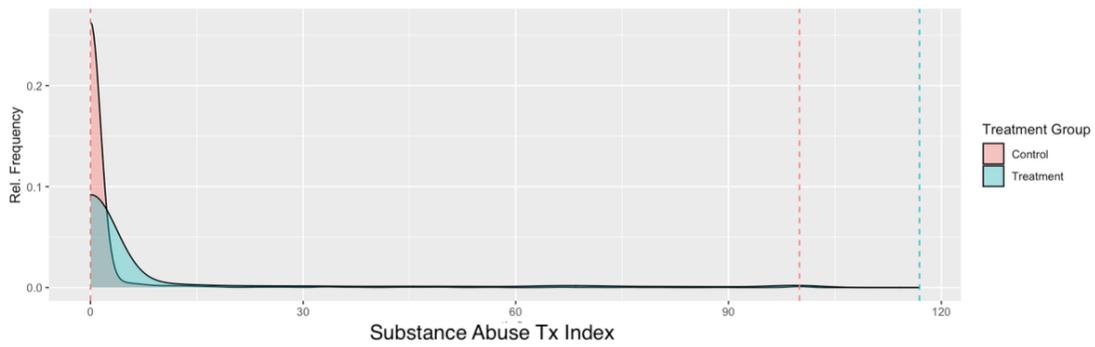

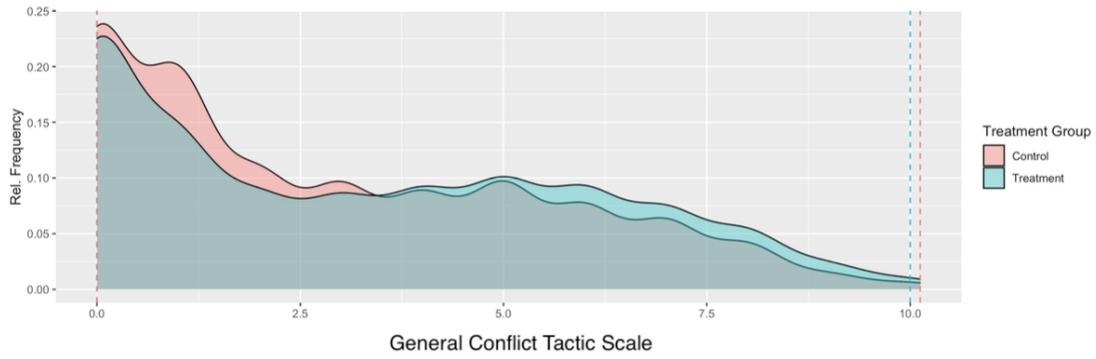
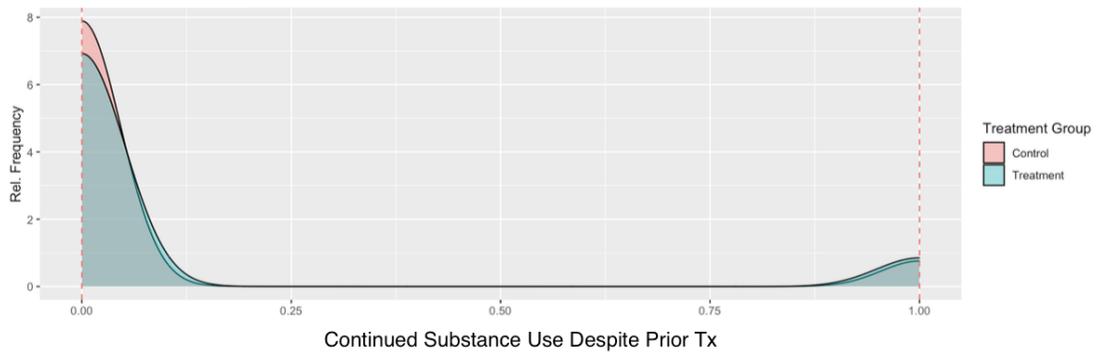
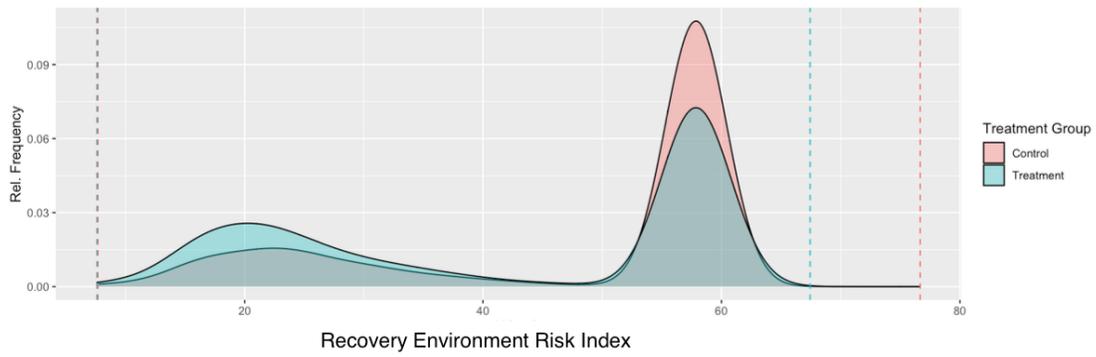
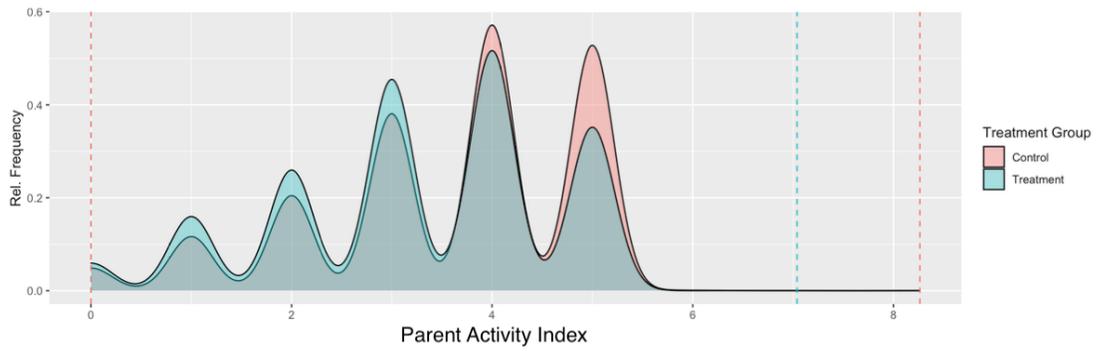

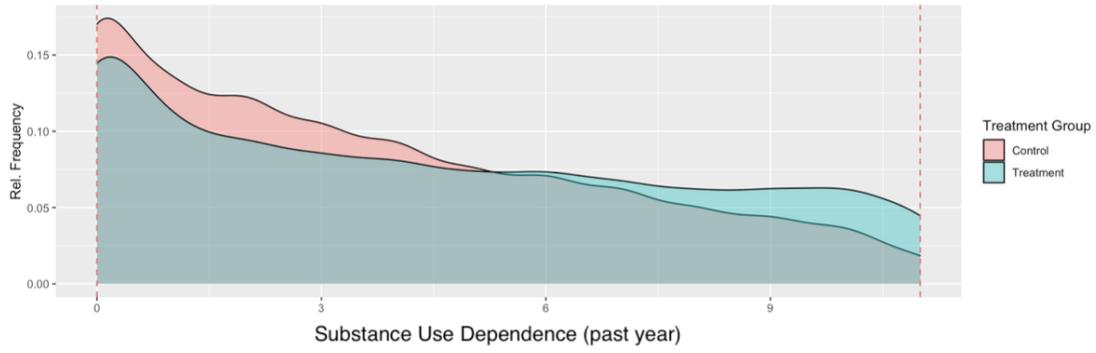
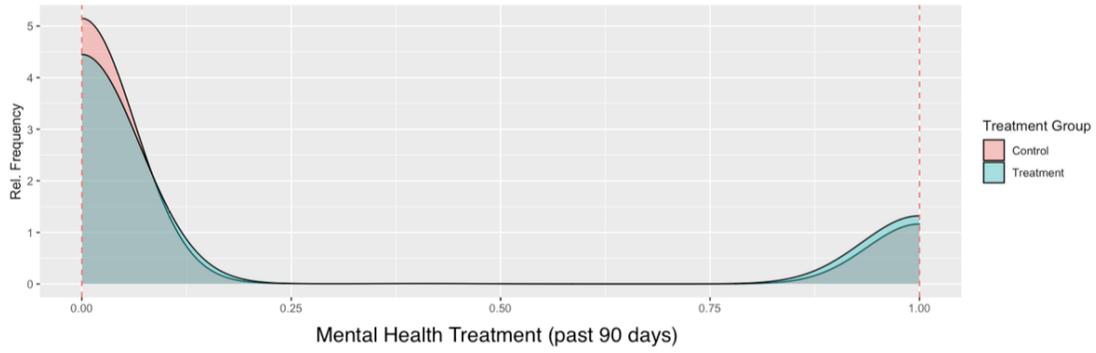
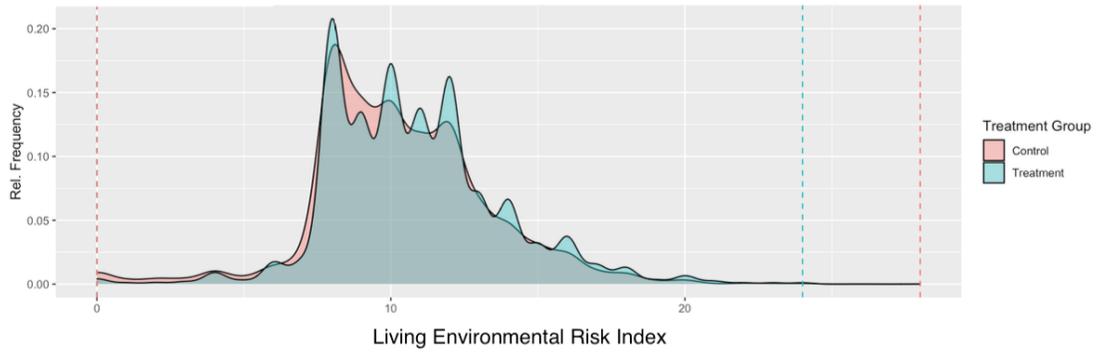
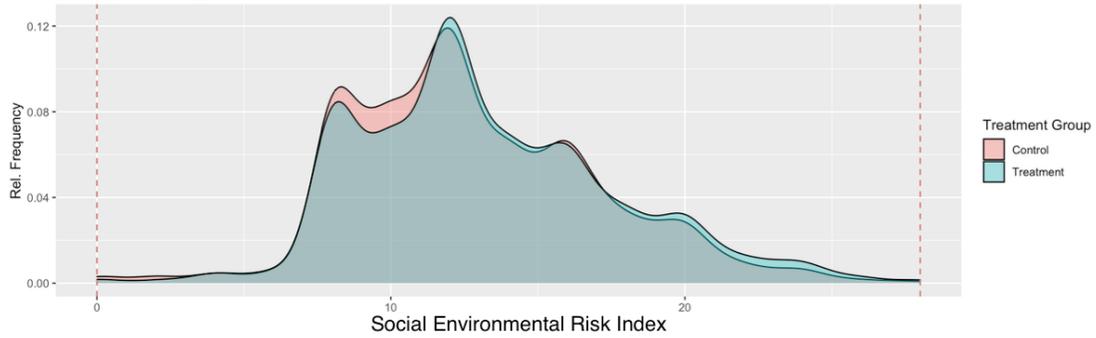